\documentclass[onecolumn,showpacs,amsmath,amssymb,amsfonts,a4paper,aps,prd,10pt]{revtex4}
\usepackage{graphicx}
\usepackage{epstopdf}
\newcommand{\f}[2]{\frac{#1}{#2}}
\newcommand{\mc}[1]{\mathcal{#1}}
\def\p{\prime}

\def\mn{_{\mu \nu}}

\def\Mp{M_{pl}}
\begin{document}
\title{Curvature perturbations of Quasi-Dilaton non-linear massive gravity}
\author{Zahra Haghani$^2$}
\email{z\_haghani@sbu.ac.ir}
\author{Hamid Reza Sepangi$^{2}$}
\email{hr-sepangi@sbu.ac.ir}
\author{Shahab Shahidi$^{2}$}
\email{s\_shahidi@sbu.ac.ir}
\affiliation{$^2$Department of
Physics, Shahid Beheshti University, G. C., Evin, Tehran 19839,
Iran}
\begin{abstract}
We study the cosmological perturbations of the recently proposed extension of  non-linear massive gravity with a scalar field. The added scalar field ensures a new symmetry on the field space of the theory. The theory has the property of having a flat dS solution, in contrast to the standard dRGT massive gravity. The tensor part is the same as that of the standard dRGT and shows gravitational waves with a time  dependent mass which modifies the dispersion relation. We obtain the curvature perturbation of the model on superhorizon scales for a specific choice of $\omega=0$  and find that the theory does not allow a constant curvature perturbation on the superhorizon scales and we have always a growing mode. The consistency of equations restricts the parameter space of the theory.
\end{abstract}
\maketitle
\section{Introduction}
Gravity, as  the oldest  force known to man has also had the longest history of trials and tribulations along the road to discovering its nature. Over the course of its development, it has
been witnessing a myriad of attempts to unlock its notoriously difficult and mysterious behavior
from the largest to the smallest of distances. The challenge has been truly spectacular. Since
the first formulation of the gravitational field by Newton and  centuries later by Einstein in
the form of the theory of general relativity (GR), the scenery is still cluttered with debris left from
various attempts to understand its hard to grasp nature. Even today, the challenge is as fresh and
as interesting as ever. Not surprisingly, building on GR, the last couple of decades have been particularly rich in new ideas and approaches which attempt to formulate the gravitational field in such a way as to pave the way to a  formulation of the theory which would explain such recently observed phenomenon as the accelerated expansion of the universe, galaxy rotation curves and even the birth of the universe. One such attempt has been surfacing over the past few years in the form of what is now known as massive gravity which, as the name suggests, is a theory with a massive graviton as the building block of the gravitational field.

The notion of a massive  graviton has been a tempting and challenging premise in theoretical physics. One of the main motivations of having a massive graviton is that gravity  could become weak at large distances, thus  mimicking the effect of accelerated expanding  universe. The problem is not as easy as it seems. The first attempt to build a theory for a massive spin-2 field was back in 1939 when Fierz and Pauli (FP) developed a linear theory for a massive graviton \cite{FP}. It took thirty years for physicists to find out that the theory does not reduce to the standard GR when one takes the limit $m\rightarrow0$ \cite{VDVZ}. The problem was soon addressed by Vainstein \cite{Vain} who proposed that adding non-linearities to the action can cure the problem and screen the effect of helicity-0 component of the massive graviton at solar system scales. The simplest possible non-linearity one can add to the FP action is by replacing the linear kinetic term for the helicity-2 field with the fully non-linear, and
still ghost free, Einstein-Hilbert action. However, the
resulting action have proven to have a ghost instability which was discovered by Boulware and Deser \cite{BD}. The problem arises because the lapse function is no longer a Lagrange multiplier. This new problem can then be solved if one appropriately adds interaction terms to the Lagrangian and again makes the lapse function a Lagrange multiplier order by order in non-linearities \cite{arkani,cremine}.

In this paper, after a brief review of the theory we study the curvature perturbation around an accelerating solution and obtain the background equations. The second order Lagrangian can then be obtained using the perturbed FRW metric. It is an immediate observation from the form of the second order Lagrangian that the tensor, vector and scalar modes do not couple. The tensor mode shows a massive gravitational wave with a time-dependent mass parameter which we shall obtain in section \ref{tensorsec}. In section \ref{vectorsec} we consider the vector mode and show that the vector part of the action vanishes at superhorizon scales and subsequently obtain the scalar mode in section \ref{scalarsec}. Two of the scalar modes are non-dynamical and can be integrated out immediately with three degrees of freedom remaining in the scalar sector. We show that in the superhorizon limit where one of the degrees of freedom does not play a role, the curvature perturbation can be obtained analytically and there is a vast region in the parameter space  over which the curvature perturbation will grow on the superhorizon scales.

\section{A brief review}
Recently, a two parameter theory of massive gravity has been proposed \cite{dRGT}. With the aid of a new re-summation of the non-linear interaction \cite{dRG} de Rham, Gabadadze and Tolley (dRGT) have constructed a theory which is free of ghost instabilities in the decoupling limit. The theory has been proven to be ghost-free in the full non-linear theory and hence a reliable effective field theory of a massive spin-2 field \cite{mir}. In all fairness, there are also criticisms towards dRGT in that superluminal shock wave solutions have been shown to appear in the theory \cite{deser}. However, subsequently it was shown that such shock waves are unstable with an arbitrary fast decaying time \cite{jcap}.
The Lagrangian for dRGT non-linear massive gravity can be written as
\begin{align}\label{1}
\mathcal{L}=  \frac{\Mp^2}{2} \sqrt{-g}  \bigg( R+m^2\mathcal{U}(\mathcal{K}) \bigg)+ \sqrt{- g} \mathcal{L}_m(g_{\mn},\psi) \,,
\end{align}
where the non-linear interactions are collected in $\mathcal{U}$
\begin{align}\label{2}
\mathcal U(\mathcal{K})=\mathcal{U}_2+\alpha_3\ \mathcal{U}_3+\alpha_4\ \mathcal{U}_4,
\end{align}
which consists of polynomials of various traces of the matrix $\mathcal{K}^\mu_\nu(g,\phi^a)=\delta^\mu_\nu-\sqrt{g^{\mu\alpha}f_{\alpha\nu}}$ where the fiducial metric is defined as $f_{\alpha\nu}=\partial_\alpha \phi^a
\partial_\nu \phi^b \eta_{ab}$ and $ \phi^a$ are the Stuckelberg fields responsible for the breaking of general covariance
\begin{subequations}\label{eq3}
\begin{align}\label{3}
\mathcal{U}_2&=[\mathcal{K}]^2-[\mathcal{K}^2]\,,\\
\mathcal{U}_3&=[\mathcal{K}]^3-3 [\mathcal{K}][\mathcal{K}^2]+2[\mathcal{K}^3]\,,\\
\mathcal{U}_4&=[\mathcal{K}]^4-6[\mathcal{K}^2][\mathcal{K}]^2+8[\mathcal{K}^3]
[\mathcal{K}]+3[\mathcal{K}^2]^2-6[\mathcal{K}^4],
\end{align}
\end{subequations}
where the rectangular brackets denote
traces, $[\mathcal{K}]\equiv {\rm Tr} (\mathcal{K})= \mathcal{K}^\mu_\mu$.
The first term concides with the Fierz-Pauli mass term at the linear level, and the last two terms are non-linear interactions which ensure that the theory has no ghost.

One of the interesting properties of the theory is that if one assumes the Stuckelberg fields to be in the unitary gauge defined as $\phi^a=\delta^a_\mu x^\mu$, the theory does not have a non-trivial flat FRW solution \cite{amico}. However, the theory has an open FRW solution with an additional consideration in that one must transform the field space to the open slicing of the Minkowski metric \cite{gumru1}. The cosmological perturbations around such a solution was also considered in \cite{mukohyama} where the authors found that the scalar, vector and tensor modes actually decouple and as a result the vector mode does not play any role in the theory. The tensor mode describes a massive gravitational wave with a time-dependent mass. Cosmological evidences of dRGT theory is also considered in \cite{cosmo}. For a review on the theoretical aspect of the theory see \cite{hinter}.

One can also let the non-dynamical metric of the theory to have a kinetic term and hence construct a bimetric theory \cite{bimet1} which has been proven to be ghost free \cite{bimetr1}. The cosmological aspects of such bimetric theory is investigated in \cite{cosbi}. One may continue the procedure to build a multimetric gravity theory with dRGT non-linear mass terms. In \cite{mult1}, the authors show that the theory is ghost free in the metric formulation only when the interactions between gravitons are not cyclic. However, in \cite{mult2} the authors show that in the vielbein formulation of the theory any interactions are allowed. Also in \cite{khos} the authors show that the problem of having no flat FRW solution persists in  multimetric theories. In fact if one of the metrics is assumed to be flat, all of the other metrics will become flat.

One of the solutions to the problem of the non-existence of the flat FRW solution in the theory is by extending the theory in such a way that the graviton mass becomes a function of some scalar field $\varphi$ \cite{masva}. The cosmological solutions and dynamical analysis of such theories are considered in \cite{varcos}. Another way to extend the theory is to couple a scalar field to the mass Lagrangian such that the resulting new Lagrangian has an extra symmetry. In particular, one can couple the scalar field to the Lagrangian to achieve dilation invariance on the field space \cite{quasi}
\begin{align}\label{5}
\sigma\rightarrow\sigma-\Mp\alpha, \qquad \phi^a\rightarrow\textmd{e}^{\alpha}\phi^a.
\end{align}
In the Einstein frame one can write the new Lagrangian as
\begin{align}\label{3.1}
\mathcal{L}=  \frac{\Mp^2}{2} \sqrt{-g}  \bigg[ R -\frac{\omega}{\Mp^2} g^{\mn}\partial_\mu\sigma\partial_\nu\sigma+m^2\mathcal{U}(\tilde{\mathcal{K}}) \bigg]+ \sqrt{- g} \mathcal{L}_m(g_{\mn},\psi) \,,
\end{align}
where $\tilde{\mathcal{K}}^\mu_\nu$ is now defined as
\begin{align}\label{4}
\tilde{\mathcal{K}}^\mu_\nu(g,\phi^a)=\delta^\mu_\nu-\textmd{e}^{\sigma/\Mp}\sqrt{g^{\mu\alpha}f_{\alpha\nu}}
\end{align}
Only the pure geometric part of the above action is invariant under  transformation \eqref{5}, thus the acronym  Quasi-Dilaton (QD)  for the scalar field \cite{quasi}. This theory has been proven to be free of ghost in the Minkowski background if $\omega>6$. The most interesting feature of this theory is that it admits a flat de Sitter solution even if the Stuckelberg fields are in the unitary gauge \cite{quasi}. In our notation, the de Sitter solution is also stable in the decoupling limit for
\begin{align}\label{5.5}
\alpha_3\neq0,\quad 0<\alpha_4<\f{\alpha_3^2}{2},\quad 0\leq\omega<6.
\end{align}

\section{The Background equation}
Let us assume that the background metric is of the form
\begin{align}\label{6}
ds^2=-N(t)^2dt^2+a(t)^2\big(dx^2+dy^2+dz^2\big),
\end{align}
and the Stuckelberg fields take the form
$$\phi^0=f(t),\qquad \phi^i=\delta^i_\mu x^\mu.$$
Note that we will work in the unitary gauge. However, in order to obtain the Stuckelberg equation from the action we assume the above form for the Stuckelberg fields, and finally set $f(t)=t$. We also assume that the QD field only depends on $t$. By varying the action \eqref{1} with respect to $f(t)$, one  obtains the constraint equation as
\begin{align}\label{7}
-9m^2\Mp^2 e^{\sigma/\Mp}\left(a-e^{\sigma/\Mp}\right)\left[\f{4}{3}\alpha_4e^{2\sigma/\Mp}-\left(\f{8}{3}\alpha_4-
\alpha_3\right)ae^{\sigma/\Mp}+\left(\f{4}{3}\alpha_4+\alpha_3+\f{1}{3}\right)a^2\right]=k_1,
\end{align}
where $k_1$ is an integration constant.
We are interested in the set of equations for which $k_1=0$. In this case one can  solve the above equation by using the ansatz
$$e^{\sigma/\Mp}=Xa,$$
in the equation which results in
\begin{align}\label{8}
X=\f{3\alpha_3+8\alpha_4\pm\sqrt{9\alpha_3^2-16\alpha_4}}{8\alpha_4}, \qquad X=1.
\end{align}
The solution $X=1$ is not acceptable because in this case the effective cosmological constant vanishes, and the consistency of the theory does not allow one to have a flat cosmological solution \cite{quasi}. Putting the above ansatz in the mass part of the Lagrangian,  one can write the  Friedman and Raychaudhuri equations as
\begin{align}
&3H^2-\f{\omega}{\Mp^2}\f{\dot{\sigma}^2}{2N^2}=3M_f,\label{9}\\
&2\f{\dot{H}}{N}+3H^2+\f{\omega}{\Mp^2}\f{\dot{\sigma}^2}{2N^2}=3M_g,\label{10}
\end{align}
where we define $H=\dot{a}/(aN)$ and
\begin{align}\label{10.05}
M_f&=\frac{m^2}{16 \alpha _4^2} \bigg[3 \alpha _3^2 \left(3 (X-1) \alpha _3-1\right)+4 \left(1-4 (X-1) \alpha _3\right) \alpha _4\bigg],\\
M_g&=\frac{m^2}{3} \bigg[-6-X (-6-3 r+X+2 r X)+3 (X-1) (4+(-2+r (X-3)) X) \alpha _3+12 (X-1)^2 (r X-1) \alpha _4\bigg],
\end{align}
and $r\equiv r(t)=a(t)/N(t)$.
We will use these forms of the background equations in the second order Lagrangian. Substituting equation \eqref{8} in \eqref{9} leads to a constant Hubble parameter, showing a de Sitter solution
\begin{align}\label{10.1}
H^2\equiv H_0^2=\frac{6m^2 (1-X) \big[2+4 \alpha_3+4 \alpha_4+X (-1+(X-5) \alpha_3+4 (X-2) \alpha_4)\big]}{\omega -6 }.
\end{align}
From equation \eqref{10} we obtain that $r$ should be a constant given by
\begin{align}\label{11}
r=\frac{12 (1+2 \alpha_3+2 \alpha_4) \omega +4 X^2 (1+6 \alpha_3+12 \alpha_4) (3+\omega )-3 X^3 (\alpha_3+4 \alpha_4) (6+\omega )-3 X (1+3 \alpha_3+4 \alpha_4) (6+5 \omega )}{X (\omega-6)\big(3+9 \alpha_3+12 \alpha_4+3 X^2 (\alpha_3+4 \alpha_4)-2 X (1+6 \alpha_3+12 \alpha_4)\big)}.
\end{align}
The scalar field equation is satisfied automatically by plugging in equations \eqref{8}, \eqref{10.1} and \eqref{11}.

\section{Second order Lagrangian}
In this section we study perturbations signified by

\begin{align}
 ds^2=-N(t)^2\big[1+2\phi(t,x,y,z)\big]dt^2+2N(t)a(t)\beta_i (t,x,y,z)dtdx^i +a(t)^2\big[\delta_{ij}+h_{ij}(t,x,y,z)\big]dx^i dx^j,
\end{align}
where $\phi$, $\beta_i$ and $h_{ij}$ are the perturbation variables of the FRW metric. The perturbation of the Stuckelberg fields in unitary gauge is
\begin{align}
 \phi^a=x^a+\pi^a(t,x,y,z),
\end{align}
and the perturbation of the dilaton field about the background solution has the form
\begin{align}
 \sigma=\sigma_0(t)+\zeta(t,x,y,z).
\end{align}
Now, we consider the infinitesimal coordinate transformation
\begin{align}
 x^\mu \rightarrow x^\mu+\xi^\mu(t,x,y,z),
\end{align}
which leads to the change of the perturbed quantities  as
\begin{subequations}
\begin{align}
 \phi& \rightarrow \phi-\f{1}{N}\partial_t(N\xi^0),\\
 \beta_i& \rightarrow \beta_i+\f{N}{a} \partial_i\xi^0-\f{a}{N}\dot{\xi}_i,\\
 h_{ij} &\rightarrow h_{ij}-\partial_i\xi_j  - \partial_j \xi_i-2NH\xi^0\delta_{ij},\\
 \pi^a& \rightarrow \pi^a-\xi^a,\\
 \zeta& \rightarrow \zeta-\dot{\sigma}_0 \xi^0,
\end{align}
\end{subequations}
where a dot denotes the time derivation.

Using the perturbations of the Stuckelberg fields one can construct the gauge invariant quantities using the perturbed Stuckelberg fields in the following manner
\begin{subequations}
 \begin{align}
  \Phi &=\phi-\f{1}{N}\partial_t(N\pi^0) ,\\
  \mc{B}_i&=\beta_i+\f{N}{a}\partial_i\pi^0-\f{a}{N}\dot{\pi}_i,\\
  \mc{H}_{ij}&=h_{ij}-\partial_i\pi_j-\partial_j\pi_i-2NH\pi^0\delta_{ij},\\
  \mc{Z}&=\zeta-\dot{\sigma}_0 \pi^0.
 \end{align}
\end{subequations}
We may decompose the gauge invariant vector and tensor parts as
\begin{subequations}
 \begin{align}
\mc{B}_i&=\partial_i \beta+S_i,\\
\mc{H}_{ij}&=2\psi\delta_{ij}+\partial_i\partial_j E+\f{1}{2}(\partial_i F_j+\partial_j F_i)+\gamma_{ij},
 \end{align}
\end{subequations}
where
\begin{align}
& \partial^i S_i=0=\partial^i F_i,\nonumber\\
&\partial^i \gamma_{ij}=0=\delta^{ij}\gamma_{ij}.
\end{align}
We note that there are four degrees of gauge freedom because of the coordinate transformation, two of which represent the scalar part and the others relate to the vector part. One may fix the gauge freedom by the choice
\begin{align}
 \pi^0=0, \qquad \pi^i=0.
\end{align}
Note that with this gauge fixing all the gauge invariant perturbation variables become equal to the original one, making our calculations simpler.
Also note that the above gauge fixing is similar to the use of the unitary gauge, and so the fiducial metric takes the form
\begin{align}
 f_{\mu\nu}=\eta_{\mu\nu}.
\end{align}
The components of the $f^\mu_\nu=g^{\mu\rho}f_{\rho\nu}$ matrix in the unitary gauge are
\begin{subequations}
 \begin{align}
 f^0_0&=\f{1}{N^2}\left(1-2\phi+ 4\phi^2-\beta_i\beta^i\right)+\mc{O}(\epsilon^3),\\
f^0_i&=\f{1}{Na}\left(\beta_i-\beta^j h_{ji}-2\phi\beta_i\right)+\mc{O}(\epsilon^3),\\
f^i_0&=-\f{1}{Na}\left(\beta^i-\beta_j h^{ji}-2\phi\beta^i\right)+\mc{O}(\epsilon^3),\\
 f^i_j&=\f{1}{a^2}\left(\delta^i_j-h^i_j-\beta^i\beta_j+h^{ik}h_{kj}\right)+\mc{O}(\epsilon^3),
 \end{align}
\end{subequations}
where $\epsilon$ represents a generic perturbation parameter. To compute the components of the $\tilde{\mc{K}}^\mu_\nu$, we use the method presented in \cite{mukohyama} to expand the square root in \eqref{4}.
To zeroth order perturbation we find
\begin{align}
 \tilde{\mc{K}}^{(0)0}_{~~~~0}=1-\f{\Delta}{N},\qquad \tilde{\mc{K}}^{(0)i}_{~~~~0}=0=\tilde{\mc{K}}^{(0)0}_{~~~~i}, \qquad \tilde{\mc{K}}^{(0)i}_{~~~~j}=\left(1-\f{\Delta}{a}\right)\delta^i_j,
\end{align}
where $\Delta=e^{\sigma_0/ \Mp}$. The first and second orders are

 \begin{align}
   \tilde{\mc{K}}^{(1)0}_{~~~0}=\f{\Delta}{N}\left(\phi-\f{\zeta}{\Mp}\right),\qquad  \tilde{\mc{K}}^{(1)0}_{~~~i}=-\f{\Delta}{N(1+r)}\beta_i,\qquad  \tilde{\mc{K}}^{(1)i}_{~~~0}=\f{\Delta}{N(1+r)}\beta^i,\qquad  \tilde{\mc{K}}^{(1)i}_{~~~j}=\f{\Delta}{a}\left(\f{1}{2}h^i_j-\f{\zeta}{\Mp}\delta^i_j\right),
 \end{align}
 \begin{subequations}
 \begin{align}
  \tilde{\mc{K}}^{(2)0}_{~~~0}&=\f{\Delta}{N}\left(\f{r(r+2)}{2(r+1)^2}\beta^i\beta_i-\f{3}{2}\phi^2+\f{1}{\Mp}\zeta\phi-\f{1}{2\Mp^2}\zeta^2\right),\\
  \tilde{\mc{K}}^{(2)0}_{~~~i}&=\f{\Delta}{N(r+1)}\left(\f{r+2}{r+1}\phi \beta_i+\f{2r+1}{2(r+1)}\beta^j h_{ji}-\f{1}{\Mp}\zeta \beta_i\right),\\
  \tilde{\mc{K}}^{(2)i}_{~~~0}&=-\f{\Delta}{N(r+1)}\left(\f{r+2}{r+1}\phi \beta^i+\f{2r+1}{2(r+1)}\beta_j h^{ji}-\f{1}{\Mp}\zeta \beta^i\right),\\
  \tilde{\mc{K}}^{(2)i}_{~~~j}&=\f{\Delta}{2a}\left(\f{2r+1}{(r+1)^2}\beta^i \beta_j-\f{3}{4}h^{ik}h_{kj}
  +\f{1}{\Mp}\zeta h^i_j-\f{1}{\Mp^2}\zeta^2 \delta^i_j\right),
 \end{align}
\end{subequations}
where $r=\f{a}{N}$, as that in the background. The traces of $ \tilde{\mc{K}}$ for zero and first order perturbation are given by
\begin{align}
 [\tilde{\mc{K}}^n]^{(0)}=3(1-X)^n+(1-rX)^n,
\end{align}
\begin{align}
 [\tilde{\mc{K}}^n]^{(1)}=nrX(1-rX)^{n-1}\left(\phi-\f{1}{\Mp}\zeta\right)+\f{n}{2}X(1-X)^{n-1}\left(h-\f{6}{\Mp}\zeta\right),
\end{align}
where $X=\Delta/a$. To second order we obtain
\begin{subequations}
 \begin{align}
   [\tilde{\mc{K}}]^{(2)}&=\f{r_2}{2r_1}X\beta^i\beta_i-\f{3}{8}Xh^{ij}h_{ij}-\f{1}{2}rX\left(3\phi^2+\f{1}{\Mp^2}\zeta^2\right)+\f{1}{2\Mp}X\zeta(h+2\phi)-\f{3}{2\Mp^2}X\zeta^2,\\
    [\tilde{\mc{K}}^2]^{(2)}&=\f{r_2-Xr_3}{r_1}X\beta^i\beta_i+(4rX-3)Xr\phi^2+\f{2}{\Mp}(1-2rX)Xr\phi\zeta+\f{1}{\Mp^2}\big[3(2X-1)+r(2rX-1)\big]X\zeta^2\nonumber\\
    &+\f{1}{\Mp}(1-2X)Xh\zeta+(X-\f{3}{4})Xh^{ij}h_{ij},\\
     [\tilde{\mc{K}}^3]^{(2)}&	 =\f{3}{2r_1}(r_2-2r_3X+r_4X^2)X\beta^i\beta_i-\f{3}{8}(3-5X)(1-X)Xh_{ij}h^{ij}-\f{3}{2}(1-rX)(3-5rX)Xr\phi^2\nonumber\\
     &+\f{3}{\Mp}(1-rX)(1-3rX)Xr\phi\zeta+\f{3}{2\Mp}(1-X)(1-3X)X h\zeta\nonumber\\&
     -\f{3}{2\Mp^2}\big[(r+3)-4X(r^2+3)+3X^2(r^3+3)\big]X\zeta^2,\\
       [\tilde{\mc{K}}^4]^{(2)}&=\f{2}{r_1}\left(r_2-3Xr_3+3X^2r_4-X^3r_5\right)X\beta_i \beta^i+\f{3}{2}(2X-1)(1-X)^2Xh_{ij}h^{ij}+6(2rX-1)(1-rX)^2Xr\phi^2\nonumber\\
       &+\f{4}{\Mp}(1-4rX)(1-rX)^2Xr\phi\zeta+\f{2}{\Mp^2}\left(-r-3+6X(r^2+3)-9X^2(r^3+3)+4X^3(r^4+3)\right)X\zeta^2 \nonumber\\
       &+\f{2}{\Mp}(1-4X)(1-X)^2Xh\zeta.
 \end{align}
\end{subequations}
where
\begin{align}
r_n=\sum^{n}_{i=0}r^i.
\end{align}
From the above formulae, one may construct the mass term using equations \eqref{eq3}.
The gauge invariant second order Lagrangian can then be written as
\begin{align}\label{ac1}
S^{(2)}=\Mp^2\int d^4 x Na^3\Bigg(\mathcal{L}+\f{3}{2}M_f\left(-\Phi^2+\mc{B}^i\mc{B}_i+\Phi\mc{H}\right)+\f{3}{8}(2M_f-M_g)(\mc{H}^2-2\mc{H}_{ij}\mc{H}^{ij})+\mc{L}_{mass}\Bigg),
\end{align}
where we have defined
\begin{align}
\mc{L}&=\f{1}{8N^2}(\dot{\mc{H}}_{ij}\dot{\mc{H}}^{ij}-\dot{\mc{H}}^2)+\f{H}{N}\Phi\dot{\mc{H}}-\f{1}{a}\left(2H\Phi-\f{1}{2N}\dot{\mc{H}}\right)\partial_i\mc{B}^i-\f{1}{2Na}\partial_i\mc{B}_j\dot{\mc{H}}^{ij}-3H^2\Phi^2+\f{1}{4a^2}\left(\partial_i\mc{B}_j\partial^i\mc{B}^j-(\partial_i\mc{B}^i)^2\right)\nonumber\\&+\f{1}{2a^2}\left(\partial_i\partial_j\mc{H}^{ij}-\nabla^2\mc{H}\right)\Phi+\f{1}{8a^2}\left(2\partial^i\mc{H}_{ik}\partial_j\mc{H}^{jk}+\mc{H}_{ij}\nabla^2\mc{H}^{ij}+2\mc{H}\partial_i\partial_j\mc{H}^{ij}
-\mc{H}\nabla^2\mc{H}\right)
\nonumber\\&+\f{\omega}{\Mp^2}\left(\f{\mc{Z}^2}{2N^2}+\f{\dot{\sigma}_0}{2N^2}(\mc{H}-2\Phi)\dot{\mc{Z}}+\f{1}{2a^2}\mc{Z}\nabla^2\mc{Z}+\f{\dot{\sigma}_0}{aN}\mc{Z}\partial_i\mc{B}^i+\f{\dot{\sigma}_0^{2}}{2N^2}\Phi^2\right),
\end{align}
and
\begin{align}
\mc{L}_{mass}=M_1\mc{H}_{ij}\mc{H}^{ij}+M_2\mc{H}^2+M_{\zeta}\mc{Z}^2+(M_{h\zeta}\mc{H}+M_{\zeta\phi}\Phi)\mc{Z}+M_{\phi}\Phi^2
+M_\beta\mc{B}_i\mc{B}^i+M_{h\phi}\mc{H}\Phi.
\end{align}
The definition of $M_i$'s are given in Appendix \ref{app1}. We have also used equations \eqref{9}  and \eqref{10} to simplify the action.
The above action ensures that the scalar, vector and tensor modes do not couple to each other. Therefore we study them separately.
\subsection{Tensor mode}\label{tensorsec}
Keeping only $\gamma_{ij}$ in the action \eqref{ac1}, we obtain
\begin{align}
S^{(2)}_{tensor}=\Mp^2\int d^4x Na^3\Bigg[\f{1}{8N^2}\dot{\gamma}_{ij}\dot{\gamma}^{ ij}+\f{1}{8a^2}\gamma_{ij}\nabla^2\gamma^{ij}-\f{1}{4}(3M_g-4M_1)\gamma_{ij}\gamma^{ij}\Bigg].
\end{align}
Variation of the above action with respect to $\gamma_{ij}$ leads to the equation of motion for  tensor perturbations
\begin{equation}\label{ten1}
\frac{\partial}{\partial t}\bigg(\frac{a^3}{N}\dot{\gamma}^{ij}\bigg)-Na \nabla^2 \gamma^{ij}+2(3M_g-4M_1) Na^3\gamma^{ij}=0.
\end{equation}
Fourier transforming the above equation and using the conformal time defined as
\begin{align}
d\eta=\f{N}{a}dt,
\end{align}
one can write equation \eqref{ten1} as
\begin{align}
\bar{\gamma}^{\p\p}+\bigg[\overrightarrow{k}^2-\f{a^{\p\p}}{a}+2a^2(3M_g-4M_1)\bigg]\bar{\gamma}=0,
\end{align}
where we have dropped the indices of $\gamma_{ij}$ and define $\bar{\gamma}=\f{a}{2}\gamma$. This equation shows that the graviton acquires a time-dependent mass in this background. This is in agreement with the result of \cite{mukohyama} with a different mass parameter.
\subsection{Vector mode}\label{vectorsec}
We now study the vector mode of  action \eqref{ac1}. There are two vector modes $S^i$ and $F^i$ in the action. One can write the vector part of the action as
\begin{align}\label{acvec}
S^{(2)}_{vector}=\Mp\int d^4x Na^3\bigg[-\f{1}{16N^2}\dot{F}_i\nabla^2\dot{F}^i&+\f{1}{4Na}\dot{F}_i\nabla^2S^i-
\f{1}{4a^2}S_i\nabla^2S^i\nonumber\\&+\f{1}{2}(3M_f+2M_\beta)S_iS^i+\f{1}{8}(3M_g-4M_1)F_i\nabla^2F^i\bigg].
\end{align}
One can see from the above action that the vector mode $S^i$ is an auxiliary field. Varying the action with respect to $S^i$ gives
\begin{align}
\f{1}{4Na}\nabla^2\dot{F}_i-\f{1}{2a^2}\nabla^2S_i+(3M_f+2M_\beta)S_i=0. \label{extra}
\end{align}
Going over to the Fourier space and substituting $S^i$ from the above equation into action \eqref{acvec} we obtain
\begin{align}
S^{(2)}_{vector}=\f{\Mp}{8}\int d^4x Na^3\bigg[\f{(3M_f+2M_\beta)r^2k^2}{k^2+2(3M_f+2M_\beta)a^2}\dot F^i\dot F_i-k^2(3M_g-4M_1) F_i F^i\bigg],
\end{align}
On the superhorizon scales one can see that the vector part of the action vanishes. Note that equation (\ref{extra}), expressed in Fourier space, implies  $S_i=0$ on superhorizon scales.
\subsection{Scalar mode}\label{scalarsec}
In this section we study the scalar perturbations of action \eqref{ac1}. The scalar part of the second order Lagrangian can be written as
\begin{align}\label{sca1}
\mc{L}_{scalar}&=\f{12\Mp^2 a}{N}\Bigg[\f{1}{2}\left(M_1+M_2-\f{3}{4}M_f+\f{3}{8}M_g\right)N^2a^2(\nabla^2E)^2+\f{1}{12}N^2\left(M_{h\zeta}a^2\mc{Z}-\f{3}{4}\nabla^2\Phi\right)\nabla^2E\nonumber\\
&+\f{1}{16}N^2\psi\nabla^2(\nabla^2E)+\f{1}{8}\left(M_f+\f{2}{3}M_{h\Phi}\right)N^2a^2\Phi\nabla^2E+\left(\f{1}{3}M_1+M_2+\f{1}{4}M_f-\f{1}{8}M_g\right)N^2a^2\psi\left(3\psi+\nabla^2E\right)\nonumber\\
&+\f{1}{12}Na^2H\Phi\nabla^2\dot{E}+\f{1}{2}a\left(aNH\Phi+\f{1}{3}N\nabla^2\beta-\f{1}{6}a\nabla^2\dot{E}\right)\dot{\psi}-\f{1}{4}a^2\dot{\psi}^2+\f{3}{4}\left(M_f+\f{2}{3}M_{h\Phi}\right)N^2a^2\Phi\psi\nonumber\\
&-\f{1}{6}N^2\Phi\nabla^2\psi-\f{1}{8}\left(2H^2-\f{2}{3}M_\Phi+M_f\right)a^2N^2\Phi^2-\f{1}{6}N^2aH\Phi\nabla^2\beta+\f{1}{12}M_{\zeta\Phi}N^2a^2\mc{Z}\Phi\nonumber\\
&+\f{1}{2}M_{h\zeta}N^2a^2\mc{Z}\psi-\f{1}{12}N^2\psi\nabla^2\psi+\f{1}{12}M_\zeta N^2a^2\mc{Z}^2-\f{1}{12}\left(M_\beta+\f{3}{2}M_f\right)N^2a^2\beta\nabla^2\beta\nonumber\\
&+\f{\omega}{12\Mp^2}\left(\f{1}{2}a^2\dot{\mc{Z}}^2-a^2\dot{\sigma}_0\big(\Phi-3\psi-\f{1}{2}\nabla^2E\big)\dot{\mc{Z}}+\f{1}{2}a^2\dot{\sigma}_0^2\Phi^2+\f{1}{2}N^2\mc{Z}\nabla^2\mc{Z}+\dot{\sigma}_0Na\mc{Z}\nabla^2\beta\right)
\Bigg].
\end{align}
As is seen from the equation above,  $\beta$ and $\Phi$ are non-dynamical. Transforming back to the Fourier space, one  finds their equations of motion
\begin{align}\label{sca2}
\beta=\frac{-2 M^2 H N \Phi +\omega \mc{Z} \dot{\sigma}+2 M^2 \dot{\psi}}{M^2 (2 M_\beta + 3 M_f) a N},
\end{align}
and
\begin{align}\label{sca3}
\Phi=&\f{1}{\Lambda_\phi}\bigg[\omega\bigg(k^2NH\mc{Z}-a^2(6M_f+4M_\beta)\dot{\mc{Z}}\bigg)\dot{\sigma}-6M^2NH\bigg(\f{4}{3}k^2-6M_f-4M_\beta\bigg)\dot{\psi}-M^2N\bigg(k^2a^2H\dot{E}\nonumber\\
&+a^2N\left(\f{3}{2}M_f+M_{h\phi}\right)(k^2E-6\psi)+a^2M_{\zeta\phi}N\zeta +2k^2N\psi \bigg)\bigg],
\end{align}
where we have defined
\begin{align}\label{sca4}
\Lambda_\phi=a^2M^2N^2\big(6M_f+4M_\beta\big)\big(3M_f-2M_\phi +6H^2\big)+8k^2M^2N^2H^2-a^2\omega\big(6M_f+4M_\beta\big)\dot{\sigma}.
\end{align}
It is worth mentioning that the scalar perturbations of QD massive gravity has also been addressed in the decoupling limit in \cite{quasi} where the authors argue that only one of the scalar modes can be
captured in this limit. As we can see above, we have three scalar modes in our scalar Lagrangian. However, one combination of these scalar modes should be non-dynamical due to the ghost-free nature
of the theory \cite{quasi}.

At this point we are interested in the behavior of the fields on the superhorizon scales where $k^2\rightarrow0$. After substitution of $\beta$ and $\Phi$ from equations \eqref{sca2} and \eqref{sca3} and defining the curvature perturbation on constant quasi-dilaton hypersurface as
\begin{align}\label{sca5}
\mc{R}=\psi+\f{H}{\dot{\sigma}_0}\mc{Z},
\end{align}
the Lagrangian \eqref{sca1} takes the form
\begin{align}\label{sca6}
\mc{L}_{scalar}^{k\rightarrow0}=&-\f{6\Mp^2a^2}{r^3\big[2M_\phi-3M_f+(\omega-6)H_0^2\big]}\Bigg(\f{1}{2}r^2\left(\lambda_5a^2+2r\omega H_0^2a+r^2(2M_\phi-3M_f+\omega H_0^2)\right)\dot{\psi}^2\nonumber\\
&-r^2(r\omega H_0^2+\lambda_5a)a\dot{\mc{R}}\dot{\psi}+\f{1}{2}r^2\lambda_5a^2\dot{\mc{R}}^2-\f{1}{6}\lambda_3a^4(\mc{R}-\psi)^2-r\lambda_2a^3\psi(\mc{R}-\psi)-2r^2\lambda_1a^2\psi^2\nonumber\\
&+H_0r\left[\f{1}{6}\lambda_6a^2(\mc{R}-\psi)+ra\big[\lambda_4\psi+(\Mp M_{\phi\zeta}-\omega H_0^2)\mc{R}\big]+9r^2\left(\f{2}{3}M_{h\phi}+M_f\right)\right]a\dot{\psi}\nonumber\\
&-\f{1}{2}H_0r\left[\f{1}{3}\lambda_6a(\mc{R}-\psi)+\omega r\big[(\omega-6)H_0^2+2M_\phi+2M_{h\phi}\big]\psi\right]a^2\psi
\Bigg),
\end{align}
where we define
\begin{align}\label{lam1}
\lambda_1&=\left(\f{3}{4}M_f-\f{3}{8}M_g+M_1+3M_2\right)\left(\omega-6\right)H_0^2-\f{45}{8}M_f^2-\left(3M_1+\f{9}{2}M_{h\phi}+9M_2-\f{9}{8}M_g-\f{3}{2}M_\phi\right)M_f\nonumber\\
&+\left(6M_2-\f{3}{4}M_g+2M_1\right)M_\phi-\f{3}{2}M_{h\phi}^2,\\
\lambda_2&=\f{1}{2}\omega(\omega-6)H_0^2+\bigg((\Mp M_{h\zeta}+M_\phi+M_{h\phi})\omega-6\Mp M_{h\phi}\bigg)H_0^2\nonumber\\
&-3\Mp\left((M_{h\zeta}+\f{1}{2}M_{h\phi})M_f+\f{1}{3}M_{h\phi}M_{\zeta\phi}-\f{2}{3}M_\phi M_{h\zeta}\right),\\
\lambda_3&=-3\omega H_0^4+\left((M_\phi+\Mp^2 M_\zeta +\Mp M_{\zeta\phi})\omega-6\Mp^2 M_\zeta\right)H_0^2-3\Mp^2\left(\f{1}{6}m_{\zeta\phi}^2+M_\zeta M_f-\f{2}{3}M_\phi m_\zeta\right),\\
\lambda_4&=\f{1}{2}\omega(\omega-4)H_0^2+(M_{h\phi}+M_\phi)\omega-\Mp M_{\zeta\phi},\\
\lambda_5&=\left(H_0^2-\f{1}{3}M_\phi+\f{1}{2}M_f\right)\omega,\\
\lambda_6&=-6\lambda_5+\omega\Mp M_{\zeta\phi}.
\end{align}
For $\omega=0$, the field equations at the superhorizon scales are simplified to
\begin{align}\label{eqR}
3r^2H_0MM_{\zeta\phi}\dot{\psi}-a^2\lambda_3 (\mc{R}-\psi)-3ra \lambda_2\psi=0 ,
\end{align}
\begin{align}
r^3(2M_\phi &-3M_f)\big(r\ddot{\psi}+2aH_0\dot{\psi}\big)+a^2\bigg[r^2H_0MM_{\zeta\phi}\dot{\mc{R}}-\f{1}{3}\lambda_3 a^2(\mc{R}-\psi)+2ra(2\lambda_4 H_0^2-\lambda_2)\psi \nonumber\\
&+ra(4H_0MM_{\zeta\phi}+\lambda_2)\mc{R}+r^2\left(9H_0^2(3M_f+2M_{h\phi})+4\lambda_1\right)\psi\bigg]=0,
\end{align}
which can be analytically solved for $\mc{R}$ and $\psi$. Substituting $\mc{R}$ from \eqref{eqR} into the second equation and solving the resulting equation for $\psi$ results in
\begin{align}
\psi = t^{\frac{3}{2}}\bigg(C_1t^{\f{\sqrt{9 A^2 H_0^2-32 A B}}{2 A H_0}} +C_2t^{\frac{-\sqrt{9 A^2 H_0^2-32 AB}}{2 A H_0}}\bigg),
\end{align}
where $C_1$ and $C_2$ are integration constants and we have defined
\begin{align}
A= M_f M_\zeta-\f{2}{3} M_\phi M_\zeta+\f{1}{6}M_{\zeta\phi}^2,
\end{align}
\begin{align}
B=&\f{1}{8}\bigg(8M_1+24M_h-6M_{h\phi}-3M_g-3M_f+3\big(M_{\zeta\phi}-M_{h\zeta}\big) M_{h\zeta}\bigg)H_0^2+\f{1}{96}\bigg(24 M_{h\zeta}^2 M_\phi + 90 M_f^2 M_\zeta \nonumber\\
&+24 M_{h\phi}^2 M_\zeta -
   24 M_{h\phi} M_{h\zeta} M_{\zeta\phi} - (8 M_1 - 3 M_g + 24 M_h) (4 M_\phi M_\zeta - M_{\zeta\phi}^2) \nonumber\\
&+ 6 M_f \big(-6 M_{h\zeta}^2 + (8 M_1 - 3 M_g + 24 M_h + 12 M_{h\phi} - 4 M_{\phi}) M_\zeta -
      6 M_{h\zeta} M_{\zeta\phi} + M_{\zeta\phi}^2\big)\bigg),
\end{align}
\begin{align}
\mc{R}=\psi+t^{5/2}\left(C_3t^{\f{\sqrt{9 A^2 H_0^2-32 A B}}{2 A H_0}} +C_4t^{\frac{-\sqrt{9 A^2 H_0^2-32 AB }}{2 A H_0}}\right),
\end{align}
where $C_3$ and $C_4$ are some functions of $C_1$, $C_2$ and $M_i$.

Noting that $t$ is the conformal time, a simples analysis shows that if the condition
\begin{align}
-\f{3}{2}<\f{\sqrt{9 A^2 H_0^2-32 A B}}{2 A H_0}<\f{3}{2}, \label{eq63}
\end{align}
holds, the curvature perturbation decays on superhorizon scales. On the other hand, if we have
\begin{align}
\f{\sqrt{9 A^2 H_0^2-32 A B}}{2 A H_0}=\f{3}{2}\quad\textmd{or}\quad-\f{3}{2},\label{eq64}
\end{align}
the curvature perturbation becomes constant on superhorizon scales. However, writing the above expressions in terms of $\alpha_3$ and $\alpha_4$ one can see that conditions (\ref{eq63},\,\ref{eq64}) cannot be satisfied
for $H_0>0$. The other limits imply that the curvature perturbation grows on the superhorizon scales which restricts the constants $\alpha_3$ and $\alpha_4$ to
\begin{subequations}\label{had}
\begin{align}
 \alpha_3\leq -1 \qquad &\hspace{2mm}\mbox{and} \quad \bigg(0<\alpha_4<-\f{1}{4}(1+3\alpha_3)\quad \textmd{or}\quad-\f{1}{4}(1+3\alpha_3)<\alpha_4<\f{\alpha_3^2}{2}\bigg),\\
 -1<\alpha_3<0  \qquad &\hspace{2mm}\mbox{and} \quad  0<\alpha_4<\f{\alpha_3^2}{2}.
\end{align}
In the case $\alpha_4=-\f{1}{4}(1+3\alpha_3)$ only the growing mode survives and the constant $\alpha_3$ admits the following range
\begin{align}
-\f{1}{2}<\alpha_3<-\f{1}{3}.
\end{align}
\end{subequations}
Therefore, the only possibility for the curvature perturbation in  QD massive gravity is to grow on superhorizon scales.
\section{Conclusions and final remarks}
In this paper we have studied the cosmological perturbations of the Quasi-Dilaton massive gravity. This theory is the extension of the non-linear massive gravity theory recently proposed by de Rham, Gabadadze and Tolley through a scalar field. The scalar field is coupled to the mass term in such a way that the field space of the theory admits a dilatation invariance. This new symmetry of the theory enables us to obtain flat FRW solutions. If considered without matter, the theory predicts an accelerating solution which is the effect of the graviton mass. The stability of this solutions is considered in \cite{quasi} where the authors find that the $\omega$ parameter has to have a positive value less than $6$, and the parameter $\alpha_4$ has to be less than $\alpha_3^2/2$.

The tensor mode has a different behavior as compared to that of the standard GR but similar to the gravitational waves obtained in the dRGT massive gravity theory. The gravitational waves in this theory have a non-vanishing time-dependent mass which modifies the dispersion relation of the gravitational waves. The vector mode  has the property that it vanishes on the superhorizon scales.

In order to find the scalar spectrum of the theory, one can use the gauge invariant variables and then integrate out two of the non-dynamical variables included in the metric perturbation. The equations of motion of the remaining three scalar perturbations can then be obtained by varying the resulting action. As a matter of fact, the procedure is so difficult that  one cannot solve the equations analytically. However, as long as we are interested in studying the behavior of the theory at the superhorizon scales, we can study the lagrangian over that scales. One of the scalar perturbations does not play any role in the superhorizon scales because it always comes with a wave number. The resulting superhorizon Lagrangian can subsequently be varied with respect to $\psi$ and the curvature perturbation $\mc{R}$. The equations can be solved analytically if one assumes that the quasi-dilaton field has no kinetic term. We may then obtain conditions for which the curvature perturbation grows over the superhorizon scales,
i. e. equation \eqref{had}. However, if one considers the range of $\alpha_4$ within which the solution is stable, equation \eqref{5.5}, one may reduce the allowed parameter space to that represented by relations (\ref{had})
for $\omega=0$. One should note that the above range for the parameter space would be different if one considered  a non-zero $\omega$ parameter. It is also worth noting that relations (\ref{eq63},\ref{eq64}) imply no constant curvature perturbation on superhorizon scales.

It is worth mentioning that our results in this paper are in agreement with the
work by Wands \textit{et al.} \cite{wands} where it is proved that the comoving curvature perturbation will become constant on superhorizon scales if the energy-momentum tensor of the matter is conserved. This is so  since in the context of the present work, one can write the field equations of the metric as
\begin{align}\label{ap1}
G_{\mu\nu}=T^\sigma_{\mu\nu}+m^2X_{\mu\nu},
\end{align}
where $T^\sigma_{\mu\nu}$ is the energy-momentum tensor of the dilaton field. The  $X_{\mu\nu}$  tensor is the contribution of the graviton mass term which depends on the dilaton field, the metric and Stuckelberg fields. The covariant divergence of  $X_{\mu\nu}$ is not zero in the full theory which implies that the energy-momentum tensor of the dilaton field from which the curvature perturbation is constructed is not constant in the full theory. So, one expects to have growing modes on superhorizon scales in QD massive gravity theory.

Finally, the QD massive gravity theory has the potential of producing a reasonable inflationary scenario if one adds a potential to the action or change the graviton mass to be a function of the quasi-dilaton field. This can break the dilatation invariance, which we will study in future works.

After the completion of our paper, two more works have appeared on the same subject \cite{quasi1, quasi2} where the emphasis is on the appearance of ghosts in the scalar mode on sub-horizon scales.
\acknowledgments
We would like to thank A. E. Gumrukcuoglu for useful discussions.
\appendix
\section{Constants of the second order mass Lagrangian}\label{app1}
\begin{align}
M_1=\frac{m^2}{128 \alpha_4^2} \bigg[9 r \alpha_3^2 (-1+3 (X-1) \alpha_3)+4 \left(1+3 r-3 (-3+4 r (X-1)+X) \alpha_3-18 (X-1) \alpha_3^2\right) \alpha_4\nonumber\\
 -16 (8+9 \alpha_3 -X (7+6 \alpha_3)+3 r (-2+X-4 \alpha_3+3 X \alpha_3)) \alpha_4^2-192 (r-1) (X-1) \alpha_4^3\bigg],
\end{align}
\begin{align}
M_2=\frac{m^2}{32\alpha_4} \bigg[r (1 + 6\alpha_3) (3 (X-1)\alpha_3 -1) + 4(6-4r-3X-rX + 3(4 -3X + r (4X-5))\alpha_3\alpha_4 + 48 (r-1) (X-1)\alpha_4^2 \bigg],
\end{align}
\begin{align}
M_\zeta&=\frac{1}{32 \alpha_4^2}3 m^2\bigg[9 (7 r-3) \alpha_3^2 (3 (X-1) \alpha_3 -1)+4 \big(-4+12 r+3 (3+r+7 X-19 r X) \alpha_3+18 (9 r-5) (X-1) \alpha_3^2\big) \alpha_4\nonumber\\
&+16 (2+X (14-33 \alpha_3)+39 \alpha_3+r (6-51 \alpha_3+15 X (3 \alpha_3 -2))) \alpha_4^2+384 (r-1) (X-1) \alpha_4^3\bigg],
\end{align}
\begin{align}
M_\phi=\frac{3 m^2}{32 \alpha_4^2} \left[-3 \alpha_3^2+9 (X-1) \alpha_3^3+4 \alpha_4-16 (X-1) \alpha_3 \alpha_4\right],
\end{align}
\begin{align}
M_{h\zeta}=&\frac{m^2}{32 \alpha_4^2} \bigg[27 r \alpha_3^2 (1-3 (X-1) \alpha_3)-4 \left(2+4 r-3 (-6+3 r+2 X+7 r X) \alpha_3+18 (-2+5 r) (X-1) \alpha_3^2\right) \alpha_4\nonumber\\
&-16 (8+X (2-24 \alpha_3)+30 \alpha_3+r (-2-39 \alpha_3+X (-14+33 \alpha_3))) \alpha_4^2-384 (r-1) (X-1) \alpha_4^3\bigg],
\end{align}
\begin{align}
M_{\zeta\phi}=&\frac{m^2}{16 \alpha_4^2}\bigg[27 (2 (r-1) r-1) \alpha_3^2 (3 (X-1) \alpha_3 -1)+4 \bigg(-6+8 (r-1) r+3 (9 (X-1)+2 r (2+r+7 X-7 r X)) \alpha_3\nonumber\\
&+54 r (2 r-3) (X-1) \alpha_3^2\bigg) \alpha_4+16 (4 X-6+8 r (r-1+3 X-2 r X)+9 (1+2 (r-2) r) (X-1) \alpha_3) \alpha_4^2\bigg],
\end{align}
\begin{align}
M_{h\phi}=\frac{m^2}{8\alpha_4} \bigg[(1+6 \alpha_3) (3 (X-1) \alpha_3 -1)+4 (2-3 \alpha_3+X (-4+3 \alpha_3))\alpha_4\bigg],
\end{align}
\begin{align}
M_\beta=& -\frac{m^2}{32 r_1 \alpha _4^2}\bigg[9 \alpha _3^2\bigg(r_1+6 r r_2-2 (3+4 r) r_3+(8+3 r) r_4-3 r_5\bigg)  \bigg(3 (X-1) \alpha _3-1\bigg)\nonumber\\
&+4 \bigg\{8 r r_2-(8+9 r) r_3+3 (3+r) r_4-3 r_5+3 \bigg[r_1+\bigg(-4 (X-1) r_1-2 r (3+7 X) r_2+6 r_3+13 r r_3+14 X r_3 \nonumber\\
& +17 r X r_3-13 r_4-6 r r_4-17 X r_4-6 r X r_4+6 (1+X) r_5\bigg) \alpha _3+6 (X-1) \big[10 r r_2-5 (2+3 r) r_3\nonumber\\
&+3 (5+2 r) r_4-6 r_5\big] \alpha _3^2\bigg]\bigg\}\alpha _4+16 \bigg\{r r_2 \big[-4-25 X+(75 X-78) \alpha _3\big]+r_3 \big[4+12 r+25 X+33 r X \nonumber\\
&+3 (26+43 r-25 X-42 r X) \alpha _3\big]+6 r_5 \big[1+2 X-9 (X-1) \alpha _3\big]\nonumber\\
&+3r_4 \bigg(-4-2 r-11 X-4 r X+\big[-43+18 r (X-1)+42 X\big] \alpha _3\bigg)\bigg\} \alpha _4^2\nonumber\\&+192 \big[r (5 X-4) r_2+(4+7 r-5 X-9 r X) r_3+(-7-3 r+9 X+4 r X) r_4+(3-4 X) r_5\big] \alpha _4^3\bigg].
\end{align}


\begin{thebibliography}{99}
\bibitem{FP} M. Fierz, W. Pauli, Proc. Roy. Soc. Lond. A173, 211-232 (1939).
\bibitem{VDVZ} H. van Dam, M. J. G. Veltman, Nucl. Phys. B22, 397-411 (1970); V. I. Zakharov, JETP Lett. 12, 312 (1970).
\bibitem{Vain} A. I. Vainshtein, Phys. Lett. B 39, 393 (1972).
\bibitem{BD} D. G. Boulware, S. Deser, Phys. Rev. D6, 3368-3382 (1972).
\bibitem{arkani}N. Arkani-Hamed, H. Georgi, M. D. Schwartz, Annals Phys. 305, 96-118 (2003).
\bibitem{cremine}P. Creminelli, A. Nicolis, M. Papucci, E. Trincherini, JHEP 0509, 003 (2005).
\bibitem{dRGT} C. de Rham, G. Gabadadze, A. J. Tolley, Phys. Rev. Lett. 106, 231101 (2011).
\bibitem{dRG}C. de Rham, G. Gabadadze, Phys. Rev. D82, 044020 (2010).
\bibitem{mir}S. F. Hassan and R. A. Rosen, JHEP 04, 123 (2012), arXiv:1111.2070 [hep-th]; S. F. Hassan, R. A. Rosen and A. Schmidt-May, JHEP 02, 026 (2012); S. F. Hassan and R. A. Rosen, Phys. Rev. Lett. 108, 041101 (2012), arXiv:1106.3344 [hep-th]; M. Mirbabayi, Phys. Rev. D 86, 084006 (2012).
\bibitem{deser} S. Deser and A. Waldron, Phys. Rev. Lett. 110, 111101 (2013).
\bibitem{jcap} C. Burrage, C. de Rham, L. Heisenberg and A. J. Tolley, JCAP 07, 004 (2012).
\bibitem{amico} G. D'Amico, C. de Rham, S. Dubovsky, G. Gabadadze, D. Pirtskhalava, A. J. Tolley, [arXiv:1108.5231].
\bibitem{gumru1}  A. E. Gumrukcuoglu, C. Lin and S. Mukohyama, JCAP 11, 030 (2011), [arXiv:1109.3845].
\bibitem{mukohyama} A. Emir Gumrukcuoglu, Chunshan Lin and Shinji Mukohyama, JCAP 03 (2012) 006, arXiv:1111.4107v2 [hep-th].
\bibitem{cosmo} P. Gratia, W. Hu and M. Wyman,  Phys. Rev. D 86, 061504 (2012), [arXiv:1205.4241]; G. D’Amico, Phys. Rev. D 86, 124019 (2012),
[arXiv:1206.3617];  M. Fasiello and A. J. Tolley, JCAP 11, 035 (2012), [arXiv:1206.3852]; Y. Gong, [arXiv:1207.2726]; M. S. Volkov, Phys. Rev. D 86, 104022 (2012), [arXiv:1207.3723]; C. -IChiang, K. Izumi and P. Chen, JCAP 12, 025 (2012), [arXiv:1208.1222]; H. Motohashi and T. Suyama, Phys. Rev. D 86, 081502 (2012), [arXiv:1208.3019]; D. Langlois and A. Naruko, [arXiv:1206.6810].
\bibitem{hinter}  K. Hinterbichler, Rev. Mod. Phys. 84, 671 (2012), [arXiv:1105.3735].
\bibitem{bimet1} S. F. Hassan and R. A. Rosen, JHEP 02, 126 (2012), [arXiv:1109.3515].
\bibitem{bimetr1} S. F. Hassan and R. A. Rosen, JHEP 04, 123 (2012), [arXiv:1111.2070].
\bibitem{cosbi}  N. Khosravi, H. R. Sepangi and S. Shahidi, Phys. Rev. D 86, 043517 (2012); D. Comelli, M. Crisostomi, F. Nesti and L. Pilo, JHEP 03, 067 (2012) [Erratum-ibid. 06, 020 (2012)], [arXiv:1111.1983]; M. S. Volkov, JHEP
01, 035 (2012), [arXiv:1110.6153]; M. von Strauss, A. Schmidt-May, J. Enander, E. Mortsell and S. F. Hassan, JCAP 03, 042 (2012),
[arXiv:1111.1655]; M. Crisostomi, D. Comelli and L. Pilo, JHEP 06, 085 (2012), [arXiv:1202.1986]; M. S. Volkov, Phys. Rev. D 86, 061502 (2012), [arXiv:1205.5713]; S. ’i. Nojiri and S. D. Odintsov, Phys.
Lett. B 716, 377 (2012), [arXiv:1207.5106]; Y. Akrami, T. S. Koivisto and M. Sandstad, JHEP 03, 099 (2013).
\bibitem{mult1} K. Nomura and J. Soda, Phys. Rev. D 86, 084052 (2012).
\bibitem{mult2} K. Hinterbichler, R. A. Rosen, JHEP 07 (2012) 047.
\bibitem{khos}N. Khosravi, N. Rahmanpour, H. R. Sepangi and S. Shahidi, Phys. Rev. D 85, 024049 (2012), [arXiv:1111.5346].
\bibitem{masva} Q. -G. Huang, Y. -S. Piao and S. -Y. Zhou, Phys. Rev. D 86,
124014 (2012), [arXiv:1206.5678].
\bibitem{varcos} E. N. Saridakis, [arXiv:1207.1800]; Y. -F. Cai, C. Gao and E. N. Saridakis, JCAP 10, 048 (2012), [arXiv:1207.3786]; G. Leon, J. Saavedra, E. N. Saridakis, arXiv:1301.7419 [astro-ph.CO]; K. Hinterbichler, J. Stokes and M. Trodden, [arXiv:1301.4993]; D. -J. Wu, Y. -F. Cai and Y. -S. Piao, [arXiv:1301.4326].
\bibitem{quasi} G. D'Amico, G. Gabadadze, L. Hui and D. Pirtskhalava, [arXiv:1206.4253].
\bibitem{wands} D. Wands, K. A. Malik, D. Lyth and A. R. Liddle, Phys. Rev. D 62, 043527 (2000).
\bibitem{quasi1} G. D'Amico, G. Gabadadze, L. Hui, D. Pirtskhalava, arXiv:1304.0723 [hep-th].
\bibitem{quasi2} A. E. Gumrukcuoglu, K. Hinterbichler, C. Lin, S. Mukohyama, M. Trodden, arXiv:1304.0449 [hep-th].
\end{thebibliography}
\end{document}